\documentclass[11pt]{article}
\ifdefined\XeTeXversion\else\pdfoutput=1\fi  
\usepackage[utf8]{inputenc}
\usepackage[T1]{fontenc}
\usepackage{lmodern}
\usepackage{microtype}
\usepackage{amsmath,amssymb,amsthm}
\usepackage{booktabs}
\usepackage{graphicx}
\usepackage[round]{natbib}
\usepackage[margin=1.05in]{geometry}
\usepackage[colorlinks=true,linkcolor=blue,citecolor=blue,urlcolor=blue]{hyperref}
\usepackage{url}
\newtheorem{definition}{Definition}
\newtheorem{problem}{Problem}


\title{Predicting What Answer Engines Cite:\\
Open Decision Models, Rerankers and a Distilled 4B Student\\
on Ten Engines under Controlled Retrieval\thanks{The model, the benchmark, the predictions and
the code are released; see Section~\ref{sec:repro}.}}

\author{Elisha Bajemon\\
TW3 Partners (Citead)\\
\texttt{e.bajemon@tw3partners.com}
\and
André-Louis Rochet\\
TW3 Partners (Citead)\\
\texttt{arochet@tw3partners.com}}

\date{October 2026}

\begin{document}
\maketitle

\begin{abstract}
Generative engine optimization needs a scorer that ranks candidate pages by the share of an
answer they will receive as citations. We measure pretrained rerankers, a hosted typed decision
model (Jev), eight open decision models and distilled students on GEO-Bench queries answered by
ten engines that receive the same candidate sources in their prompt. Where several engines
answer the same query, they agree with one another at a within-group Spearman correlation of
0.704. Jev and two open 27B decision models, used without training, are the strongest scorers at
0.575 to 0.578. A Qwen3-Reranker-4B fine-tuned with low-rank adapters on teacher labels for other
GEO-Bench queries, released as GEO Citation Ranker 4B, reaches 0.542 to 0.549 across three seeds,
and its differences with the 27B models and with Jev are not distinguishable from zero on 229
queries. Most of the remaining
distance between the student and Jev disappears when Jev scores one passage at a time, as the
student does. Students trained on teacher labels and on observed engine citations reach the same
level, and labels from Jev alone match the two-teacher target at 0.6B. We release the model, the
benchmark, the predictions of every reported scorer and the code.
\end{abstract}

\section{Introduction}
\label{sec:intro}

Work on generative engine optimization (GEO) asks, for a given query, which of the candidate
pages an answer engine will draw on, and how much \citep{aggarwal2024geo}. Measuring engines
directly answers it, but each measurement is paid, slow and non-deterministic, so practitioners
increasingly use a model to predict the answer before any engine is queried. Pretrained rerankers
are free to run and were trained for topical relevance
\citep{nogueira2020monot5,chen2024bgem3,zhang2025qwen3embedding}. LLM judges prompted with a
rubric can rank sources, at the cost of one generation per group \citep{zheng2023judge}. Typed
decision models, such as the hosted Jev \citep{typesafe2026jev,shea2026jev}, return a probability
distribution over a fixed scale for each question asked about a state. Several open decision
models were published on the Hugging Face Hub after Jev's release
\citep{hf_jev27b,hf_pplx,hf_neohorse,hf_decider4b,hf_opendecider,hf_kev4b,hf_laya}. We know of no
measurement of decision models or distilled rerankers against the citations of answer engines.

This paper measures them on one benchmark of 229 GEO-Bench queries \citep{geobench}. Each query is
answered by one to ten engines that receive the same five candidate sources in their prompt, and
the label of a source is its citation share in the answer. For the 150 queries answered by
several engines, the benchmark also measures how far the engines agree with each other, which
gives every scorer a reference that comes from the engines themselves.

The paper makes three contributions. First, it reports the agreement between engines and the
level of 36 scorer configurations on the same 1220 engine-query groups, with intervals clustered
by query and paired differences (Section~\ref{sec:results}). Second, it shows that two open 27B
decision models match the hosted judge without training, and that presenting the five candidates
together accounts for most of the distance between Jev and a single-passage student. Third, it
distils teacher labels into Qwen3-Reranker students at 0.6B, 4B and 8B, releases the 4B student
as GEO Citation Ranker 4B, and separates the effect of teacher choice, label volume, capacity and
observed citations.

\section{Related work}
\label{sec:related}

\paragraph{Generative engine optimization.} \citet{aggarwal2024geo} introduced GEO and GEO-Bench
and reported citation gains from content edits. Later work measured conversational SEO on its own
benchmark \citep{puerto2025cseo}, ranked the factors of competition between two sources
\citep{vishwakarma2026whatgetscited}, separated being cited from shaping the answer
\citep{zhang2026absorption}, and repaired document-specific citation failures with agents
\citep{tian2026agentgeo}. Adversarial manipulation of conversational search rankings was
demonstrated by \citet{pfrommer2024ranking}. \citet{bajemon2026scoring} re-measured the 2023
content levers on ten current engines and found that query-agnostic content scores predict
citation only weakly. This paper uses the engine answers collected for that study.

\paragraph{Citation in generated answers.} Generated answers often cite sources that do not
support them \citep{liu2023verifiability}, and citation quality can be evaluated and trained for
\citep{gao2023alce}. LLMs also use the middle of a long context less than its ends
\citep{liu2024lost}, which predicts an effect of source position in our prompts.

\paragraph{LLM rerankers and distillation.} LLMs rerank well zero-shot \citep{sun2023rankgpt},
and their rankings can be distilled into smaller open models
\citep{pradeep2023rankvicuna,pradeep2023rankzephyr} or used to generate training data
\citep{bonifacio2022inpars}, following the general recipe of knowledge distillation
\citep{hinton2015distilling}. Qwen3-Reranker frames relevance as a yes or no answer scored by
token probabilities \citep{zhang2025qwen3embedding,yang2025qwen3}. We train with a pairwise
logistic loss \citep{burges2005ranknet} and low-rank adapters \citep{hu2022lora}.

\paragraph{Judges.} LLM-as-a-judge is standard for graded criteria \citep{zheng2023judge}, and
judges are sensitive to the order in which candidates are presented \citep{wang2023fair}. Typed
decision models were introduced for binary agent-evaluation criteria \citep{shea2026jev}; we
apply them to a graded, comparative criterion.

\section{Problem statement}
\label{sec:problem}

\begin{problem}[Citation ranking]
An engine $e$ answers query $q$ from five candidate sources $s_1,\dots,s_5$ given in its prompt
in that order. A scorer sees $q$ and the source texts, never the engine's answer, and must rank
the five sources by the citation share each receives in the answer.
\end{problem}

\begin{definition}[Citation share]
The label $y_{e,q,i}$ of source $s_i$ is its position-adjusted word count share: the words of
each answer sentence citing $s_i$, weighted by the sentence's position in the answer, summed and
normalised over the five sources \citep{aggarwal2024geo,bajemon2026scoring}.
\end{definition}

\begin{definition}[Score]
For a scorer with predictions $\hat y$, the score of a group $(e,q)$ is the Spearman correlation
between $\hat y_{\cdot}$ and $y_{e,q,\cdot}$ over the five sources, with average ranks for ties
and zero for a constant prediction. The benchmark score is the mean over groups whose five labels
are not all equal.
\end{definition}

\begin{definition}[Cross-engine agreement]
For a group $(e,q)$, the cross-engine agreement is the score of the predictor
$\hat y_i = \operatorname{mean}_{e' \neq e} y_{e',q,i}$, the mean label that the other engines
answering $q$ give to source $s_i$. It is defined for queries answered by at least two engines.
\end{definition}

Cross-engine agreement uses information that no scorer has, namely other engines' answers to the
same prompt. It measures how much of a group's ranking is shared across engines. It serves as a
reference for the scorers and is not an upper bound on them.

\section{Benchmark and scorers}
\label{sec:setup}

\paragraph{Evaluation data.} The 229 queries are the GEO-Bench test queries for which
\citet{bajemon2026scoring} collected engine answers, with the prompt, decoding settings and
citation parser described in that work (temperature 0, pinned model identifiers). Each source is
represented by the first 320 words of its raw text. The ten engines are listed with their vendor
and coverage in Appendix~\ref{app:engines}. Seventy-nine queries are answered by one engine, 50 by
five, 23 by seven to nine and 77 by all ten, which gives 1220 non-degenerate groups over 1145
passages. Every engine receives the query and the five sources in its prompt, and no engine
retrieves documents. The benchmark therefore measures citation under controlled retrieval, the
setting of GEO-Bench.

\paragraph{Pretrained rerankers.} We score the NIM reranker (rerank-qa-mistral-4b),
bge-reranker-v2-m3 and Qwen3-Reranker at 0.6B, 4B and 8B with their default inference code and no
training, on the same 320-word passages.

\paragraph{Typed decision models.} A typed decision model receives a state and a set of questions
and returns, for each question, a probability distribution over the answer levels
\citep{shea2026jev}. We ask one six-level question per passage, from level 0 (not cited) to level
5 (the answer relies mainly on this passage), and the score is the expectation over levels. The
wording is given in Appendix~\ref{app:prompts}. In the listwise protocol the state holds the query
and the five passages and the five questions are asked in one call; in the pointwise protocol the
state holds the query and one passage. Passages are truncated to 250 words in the state, which is
shorter than the text engines and rerankers receive. We score the hosted Jev 1.13 (snapshot
20260917) through its API, and eight open models from the Hugging Face Hub with the runtime of
their model cards. The eight were chosen before any was scored, as the most downloaded or best
placed on their own model cards among those that accept an ordered score question. The same state
and questions are used for every model, with no prompt tuning.

\paragraph{Distilled students.} The training data are 757 groups of five sources, one for each
GEO-Bench test query outside the evaluation set that has five sources, prepared with the same
320-word recipe. None has engine labels. Two teachers label each group. Jev scores it in the
listwise protocol, for 0.10 USD over the whole set. An LLM judge produces synthetic labels for batches
of six groups, on a 0 to 100 scale, with the rubric of Appendix~\ref{app:prompts}. The two teachers
agree at a mean per-group Spearman of 0.759. The default target is the mean of the two teachers,
each rescaled to $[0,1]$; ablations use one teacher. Students are Qwen3-Reranker models at 0.6B,
4B and 8B scored by the log-odds of ``yes'' against ``no'' after an instruction asking whether an
answer engine would cite the passage. A rank-16 adapter on all attention and feed-forward
projections is trained with a pairwise logistic loss within each group, over the pairs whose
targets differ by more than 0.05 and weighted by that difference, for four epochs at learning rate
$10^{-4}$. The 4B student trains in about two hours on one RTX~3090. The 4B student trained with
seed 42 is released as GEO Citation Ranker 4B.

\section{Evaluation protocol}
\label{sec:protocol}

Intervals are 95\% percentile bootstraps over queries, so that the groups of one query are
resampled together \citep{field2007bootstrap}, with 1000 resamples and a fixed seed. Every
comparison between two scorers is a paired difference on the groups both cover, and we call two
scorers indistinguishable when the 95\% interval of their difference covers zero. Equivalence within a margin of 0.05 is
tested by two one-sided tests, which require the 90\% interval of the difference to lie inside the
margin \citep{lakens2017equivalence}. As a second metric we report top-1 accuracy, the share of
groups where the scorer's top source is a most-cited source. Because the 77 queries answered by
all ten engines contribute most groups, we also report the mean per engine and per query.

Students trained on observed citations are evaluated by five-fold cross-validation over the
evaluation queries, with a fixed assignment of queries to folds, and each group is scored only by
the model that did not see its query. Fifteen evaluation passages share their first 40 words with
a training source, because GEO-Bench reuses pages. Removing the 57 groups that contain them changes
the score of any student by at most 0.006 and lowers the 4B student by 0.003.

\section{Results}
\label{sec:results}

\begin{table}[t]
\centering
\footnotesize
\setlength{\tabcolsep}{3pt}
\begin{tabular}{@{}llccc@{}}
\toprule
Scorer & Size & $\rho$ [95\% CI] & Top-1 & GEO Citation Ranker 4B $-$ scorer \\
\midrule
\multicolumn{5}{@{}l}{\emph{Reference points}} \\
Other engines on the same query$^a$ & -- & 0.704 [0.669, 0.737] & -- & $-$0.159 [$-$0.201, $-$0.116] \\
Source position in the prompt & -- & 0.131 [0.068, 0.190] & 0.328 & -- \\
BM25 & -- & 0.131 [0.063, 0.203] & -- & -- \\
\midrule \multicolumn{5}{@{}l}{\emph{Pretrained rerankers, zero-shot}} \\
NIM rerank-qa-mistral-4b & 4B & 0.386 [0.329, 0.449] & 0.475 & +0.156 [+0.104, +0.212] \\
bge-reranker-v2-m3 & 0.6B & 0.412 [0.351, 0.472] & 0.495 & +0.130 [+0.076, +0.179] \\
Qwen3-Reranker-0.6B & 0.6B & 0.341 [0.277, 0.409] & 0.423 & +0.201 [+0.148, +0.251] \\
Qwen3-Reranker-4B & 4B & 0.452 [0.394, 0.509] & 0.476 & +0.090 [+0.059, +0.121] \\
Qwen3-Reranker-8B & 8B & 0.449 [0.394, 0.506] & 0.479 & +0.093 [+0.050, +0.134] \\
\midrule \multicolumn{5}{@{}l}{\emph{Typed decision models, zero-shot, five passages per call}} \\
Jev 1.13, hosted & n/a & 0.575 [0.523, 0.623] & 0.563 & $-$0.033 [$-$0.072, +0.006] \\
JEV-27B & 27B & 0.578 [0.529, 0.621] & 0.566 & $-$0.037 [$-$0.080, +0.013] \\
pplx-decider-v1.1-27b & 27B & 0.575 [0.524, 0.623] & 0.580 & $-$0.033 [$-$0.070, +0.005] \\
JEV-9B & 9B & 0.516 [0.463, 0.565] & 0.475 & +0.026 [$-$0.015, +0.072] \\
NeoHorse-Jev-4B$^b$ & 4B & 0.501 [0.445, 0.556] & 0.507 & +0.048 [+0.002, +0.101] \\
\midrule \multicolumn{5}{@{}l}{\emph{Distilled students (this work), one passage per call}} \\
0.6B, seed 42 & 0.6B & 0.457 [0.402, 0.512] & 0.474 & +0.085 [+0.051, +0.119] \\
GEO Citation Ranker 4B (seed 42) & 4B & 0.542 [0.491, 0.591] & 0.567 & -- \\
4B, seed 1337 & 4B & 0.546 [0.491, 0.596] & 0.563 & $-$0.004 [$-$0.023, +0.017] \\
4B, seed 7 & 4B & 0.549 [0.496, 0.598] & 0.566 & $-$0.007 [$-$0.026, +0.014] \\
4B, three-seed ensemble & 4B & 0.549 [0.497, 0.600] & 0.568 & $-$0.007 [$-$0.020, +0.008] \\
8B, seed 42 & 8B & 0.533 [0.479, 0.584] & 0.549 & +0.009 [$-$0.015, +0.032] \\
8B, seed 1337 & 8B & 0.559 [0.507, 0.608] & 0.577 & $-$0.017 [$-$0.042, +0.005] \\
 
\bottomrule
\end{tabular}
\caption{Citation ranking on 229 queries and 1220 engine-query groups. $\rho$ is the mean
within-group Spearman correlation with observed citation share; the last column is the paired
difference with GEO Citation Ranker 4B, with its 95\% query-clustered interval. Decision models are shown
with the listwise protocol; Table~\ref{tab:protocol} gives both protocols. A random choice reaches
a top-1 accuracy of 0.23. $^a$Cross-engine agreement (Definition 3) on the 1141 groups whose query
has at least two engines; its paired difference is computed on those groups. $^b$The five-passage
state exceeds the model's 2048-token context for 23 queries; its row uses the remaining 1011
groups.}
\label{tab:main}
\end{table}

\begin{figure}[t]
\centering
\includegraphics[width=0.95\linewidth]{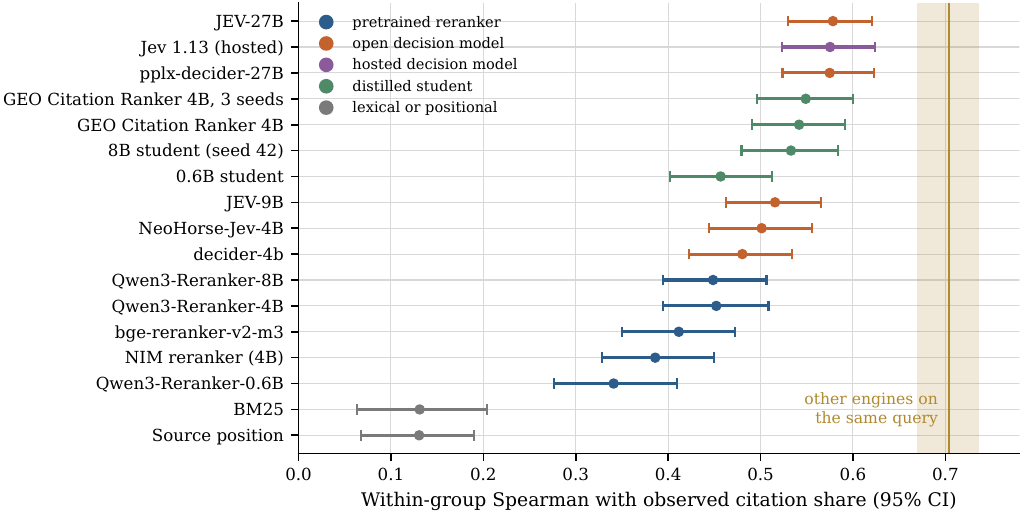}
\caption{Within-group Spearman correlation with observed citation share, with 95\%
query-clustered intervals, for a selection of the scorers of Table~\ref{tab:main}. The shaded band
is the cross-engine agreement and its interval. Decision models are shown with the listwise
protocol.}
\label{fig:scorers}
\end{figure}

\paragraph{Result 1 (engines agree more than scorers predict).} On the 1141 groups whose query has
at least two engines, the engines agree with one another at 0.704 [0.669, 0.737]
(Table~\ref{tab:main}, Figure~\ref{fig:scorers}). The agreement is 0.689 [0.652, 0.723] when
engines of the same vendor are excluded from the mean. It exceeds Jev by 0.126 [0.083, 0.175] and
the 4B student by 0.159 [0.116, 0.201] on the same groups. Engines give the first source a mean
share of 0.234 and the fifth 0.164, and ranking by position alone scores 0.131, the same as BM25.
Adding the source position to a scorer's ranking, with a weight fitted on four folds and applied
to the fifth, changes the student by +0.011 [$-$0.002, +0.025] and Jev by $-$0.013 [$-$0.023,
$-$0.005]. The distance between the scorers and the engines' agreement is therefore not explained
by source position.

\paragraph{Result 2 (open 27B decision models match the hosted judge).} Used without training,
JEV-27B and pplx-decider-27B reach 0.578 and 0.575, against 0.575 for the hosted Jev. Their paired
differences with Jev are +0.003 [$-$0.034, +0.036] and 0.000 [$-$0.028, +0.031]. Scores fall as
model size decreases: JEV-9B reaches 0.516, the four open 4B decision models 0.44 to 0.50 with
their better protocol, and a 0.4B encoder 0.214. The pretrained rerankers range from 0.341 to
0.452, and the 8B reranker is not above the 4B ($-$0.004 [$-$0.043, +0.035]).

\paragraph{Result 3 (GEO Citation Ranker 4B is not distinguishable from the 27B models).} The 4B student
scores 0.542 with seed 42, the released GEO Citation Ranker 4B, and 0.546 and 0.549 with two other
seeds, and the largest difference between seeds is
0.007, with an interval covering zero. Fine-tuning adds 0.090 [0.059, 0.121] over the same
reranker without training. Its differences with JEV-27B ($-$0.037 [$-$0.080, +0.013]),
pplx-decider-27B ($-$0.033 [$-$0.070, +0.005]), Jev ($-$0.033 [$-$0.072, +0.006]) and JEV-9B
(+0.026 [$-$0.015, +0.072]) all cover zero. Equivalence with Jev at a margin of 0.05 is not
established, since the 90\% interval of Jev's lead over the mean of the three seeds is [$-$0.002,
+0.060]. Taken one comparison at a time, the student is above each of the four open 4B decision
models, by 0.048 to 0.104. On top-1 accuracy the student (0.567), Jev (0.563), JEV-27B (0.566) and
pplx-decider-27B (0.580) are within 0.02 of one another, against 0.476 for the reranker without
training. Weighting each engine or each query equally gives the same picture with lower levels:
per query, the student scores 0.529, Jev 0.562 and JEV-27B 0.553 (Appendix~\ref{app:engines}).

\begin{table}[t]
\centering
\footnotesize
\setlength{\tabcolsep}{3pt}
\begin{tabular}{@{}llccc@{}}
\toprule
Model, zero-shot & Size & Five passages per call & One passage per call & Difference [95\% CI] \\
\midrule
Jev 1.13 (hosted) & n/a & 0.575 & 0.529 & +0.046 [+0.009, +0.087] \\
JEV-27B & 27B & 0.578 & 0.540 & +0.038 [+0.004, +0.071] \\
pplx-decider-v1.1-27b & 27B & 0.575 & 0.559 & +0.015 [$-$0.013, +0.047] \\
JEV-9B & 9B & 0.516 & 0.468 & +0.047 [$-$0.003, +0.099] \\
NeoHorse-Jev-4B & 4B & 0.501$^b$ & 0.465 & +0.049 [0.000, +0.100] \\
decider-4b & 4B & 0.480 & 0.490 & $-$0.009 [$-$0.049, +0.031] \\
opendecider-small-td & 4B & 0.486 & 0.430 & +0.056 [+0.002, +0.109] \\
kev-4b & 4B & 0.416 & 0.438 & $-$0.022 [$-$0.077, +0.039] \\
laya-typed-decisions & 0.4B & n/a$^c$ & 0.214 & -- \\
\midrule
GEO Citation Ranker 4B & 4B & -- & 0.542 & -- \\
 
\bottomrule
\end{tabular}
\caption{Decision models under the two protocols, with identical state and questions. The
difference is listwise minus pointwise on the groups both cover. $^b$1011 groups in the listwise
column (context limit). $^c$The model's 1024-token context truncates the five-passage state. The
student scores one passage per call by construction.}
\label{tab:protocol}
\end{table}

\paragraph{Result 4 (one passage at a time, the student matches Jev).} Asked one passage at a
time, Jev falls from 0.575 to 0.529 ($-$0.046 [$-$0.087, $-$0.009]) and JEV-27B from 0.578 to
0.540 ($-$0.038 [$-$0.071, $-$0.004]). Under that protocol the student's differences with Jev
(+0.012 [$-$0.026, +0.056]) and JEV-27B (+0.001 [$-$0.036, +0.040]) are close to zero. It
scores above JEV-9B (+0.074 [0.022, 0.128]). For pplx-decider-27B and the smaller decision models the
protocol effect is not distinguishable from zero (Table~\ref{tab:protocol}). Scoring the five
candidates together is the natural next step for the student.

\begin{table}[t]
\centering
\footnotesize
\setlength{\tabcolsep}{4pt}
\begin{tabular}{@{}lllc@{}}
\toprule
Student & Training labels & Training queries & $\rho$ [95\% CI] \\
\midrule
0.6B & Jev and LLM judge & 757 & 0.457 [0.402, 0.512] \\
0.6B & Jev only & 757 & 0.467 [0.417, 0.521] \\
0.6B & LLM judge only & 757 & 0.444 [0.388, 0.498] \\
0.6B & Jev and LLM judge & 189 & 0.427 [0.366, 0.485] \\
0.6B & Jev and LLM judge, cross-query distractors & 757 & 0.472 [0.416, 0.525] \\ \midrule
4B & Jev and LLM judge & 757 & 0.542 [0.491, 0.591] \\
4B & Jev and LLM judge & 189 & 0.537 [0.485, 0.587] \\
4B & observed citations, five-fold$^a$ & 183--184 & 0.541 [0.494, 0.585] \\
4B & teacher, then observed citations$^a$ & 757 + 183--184 & 0.562 [0.516, 0.610] \\
 
\bottomrule
\end{tabular}
\caption{Label ablations, all with the training recipe of Section~\ref{sec:setup} and seed 42.
$^a$Trained on the evaluation queries by five-fold cross-validation, each group scored by the
model that did not see its query; the two-stage student starts from the teacher-trained 4B
adapter.}
\label{tab:labels}
\end{table}

\paragraph{Result 5 (Jev labels and a quarter of the queries suffice).} At 0.6B, labels from Jev
alone give 0.467, labels from the LLM judge alone 0.444 and their mean 0.457, and no pairwise
difference excludes zero (Table~\ref{tab:labels}). A quarter of the training queries, 189, gives
0.537 at 4B against 0.542 with all 757 (+0.005 [$-$0.017, +0.027]); at 0.6B the difference is
0.030 [$-$0.001, +0.060]. Adding groups that mix sources from different queries changes nothing
detectable. A 4B student trained on observed citations of the evaluation queries, by five-fold
cross-validation over 183 or 184 queries per fold, reaches 0.541. At matched volume, the student
trained on teacher labels for 189 queries is not distinguishable from it ($-$0.004 [$-$0.041,
+0.028]). Fine-tuning the teacher-trained student on observed citations gives 0.562, a gain of
0.020 [$-$0.010, +0.051].

\paragraph{Result 6 (capacity stops paying at 4B).} With identical labels and recipe, the 4B
student is above the 0.6B by 0.085 [0.051, 0.119]. The 8B student is not above the 4B ($-$0.009
[$-$0.032, +0.015] at seed 42), and its two seeds differ by 0.026 [0.008, 0.046], whereas the three
4B seeds do not differ.

\section{Threats to validity}
\label{sec:threats}

\paragraph{Controlled retrieval.} Engines receive five fixed sources in their prompt, so the
benchmark measures which provided source an engine cites, and not whether a search engine
retrieves a page. This removes retrieval noise from the label. It does not establish that the
ranking transfers to engines that search the web, which would require answers collected from live
search products.

\paragraph{Teachers and inputs.} The student targets combine Jev and one LLM judge with one rubric.
At 0.6B, Jev labels alone train a student as good as the two-teacher target. Jev is a hosted model versioned by its provider, and the open 27B decision
models offer a teacher with frozen weights. Decision models see 250 words of each passage, against
320 for engines and students, which can only work against them.

\paragraph{Scope of the measurements.} The benchmark is English, built from one source dataset,
with five candidates per query, and four of the ten engines come from one vendor. Decision models
were run with one state and one set of questions, written once and not tuned. Engines change, and
the labels describe answers collected in 2026.

\section{Reproducibility}
\label{sec:repro}

\paragraph{Release.} GEO Citation Ranker 4B is released at
\url{https://huggingface.co/citead/geo-citation-ranker-4b}, with the adapters of the two other 4B
seeds and of the 0.6B student, under the Apache 2.0 license of its base model. The benchmark is
released at \url{https://huggingface.co/datasets/citead/geo-citation-benchmark}. It contains the
1145 evaluation passages, the 6345 citation labels with engine and query, the fold assignment, the
predictions of the 36 scorer configurations of this paper on every evaluation passage, the 757
training groups without labels, and the code. Passages and labels are released under CC BY-SA
4.0, the license of GEO-Bench.

\paragraph{Regimes.} The tables and the figure are regenerated offline from the released benchmark,
at no API cost. \texttt{code/reproduce\_tables.py} recomputes the correlations and paired differences,
\texttt{code/analyses.py} the cross-engine agreement, baselines, top-1 accuracy, per-engine and
per-query means and the overlap check, and \texttt{paper/make\_tables.py} and
\texttt{paper/make\_figures.py} write the tables and the figure. The agreement between the two
teachers is computed on labels that are not released. To retrain a student,
\texttt{code/label\_jev.py} labels the training groups with Jev for about 0.10 USD and
\texttt{code/train\_student.py} trains on those labels.

\paragraph{Not released.} The teacher labels are not released. The LLM-judge rubric is given in
Appendix~\ref{app:prompts}.

\paragraph{Disclosures.} The authors are employed by TW3 Partners, which offers generative engine
visibility services under the Citead name.

\section{Conclusion}
\label{sec:conclusion}

On ten engines under controlled retrieval, the strongest predictors of which source an answer
engine cites are decision models that compare the candidates together, hosted or open. GEO Citation
Ranker 4B, a 4B reranker distilled from teacher labels, is not distinguishable from them on 229
queries, and the labels it needs cost a few cents. The engines agree with one another well above every scorer we
measured, so the benchmark leaves room for scorers that compare candidates jointly. Any new scorer
can be placed against the same groups, the engines' own agreement and the same intervals.

\bibliographystyle{plainnat}
\bibliography{refs}

\begin{thebibliography}{35}
\providecommand{\natexlab}[1]{#1}
\providecommand{\url}[1]{\texttt{#1}}
\expandafter\ifx\csname urlstyle\endcsname\relax
  \providecommand{\doi}[1]{doi: #1}\else
  \providecommand{\doi}{doi: \begingroup \urlstyle{rm}\Url}\fi

\bibitem[Aggarwal et~al.(2024)Aggarwal, Murahari, Rajpurohit, Kalyan,
  Narasimhan, and Deshpande]{aggarwal2024geo}
Pranjal Aggarwal, Vishvak Murahari, Tanmay Rajpurohit, Ashwin Kalyan, Karthik
  Narasimhan, and Ameet Deshpande.
\newblock {GEO}: Generative engine optimization.
\newblock In \emph{Proceedings of the 30th ACM SIGKDD Conference on Knowledge
  Discovery and Data Mining (KDD)}, 2024.
\newblock arXiv:2311.09735.

\bibitem[{autotrust}(2026)]{hf_jev27b}
{autotrust}.
\newblock {{JEV-27B}} and {JEV-9B}.
\newblock Hugging Face, \url{https://huggingface.co/autotrust/JEV-27B}, 2026.

\bibitem[Bajemon and Rochet(2026)]{bajemon2026scoring}
Elisha Bajemon and Andr{\'e}-Louis Rochet.
\newblock Scoring without the engine: Validating a deterministic,
  manipulation-resistant content score for generative engines, end to end.
\newblock arXiv preprint, 2026.
\newblock arXiv:2609.07559.

\bibitem[Bonifacio et~al.(2022)Bonifacio, Abonizio, Fadaee, and
  Nogueira]{bonifacio2022inpars}
Luiz Bonifacio, Hugo Abonizio, Marzieh Fadaee, and Rodrigo Nogueira.
\newblock {InPars}: Data augmentation for information retrieval using large
  language models.
\newblock arXiv preprint, 2022.
\newblock arXiv:2202.05144.

\bibitem[Burges et~al.(2005)Burges, Shaked, Renshaw, Lazier, Deeds, Hamilton,
  and Hullender]{burges2005ranknet}
Chris Burges, Tal Shaked, Erin Renshaw, Ari Lazier, Matt Deeds, Nicole
  Hamilton, and Greg Hullender.
\newblock Learning to rank using gradient descent.
\newblock In \emph{Proceedings of the 22nd International Conference on Machine
  Learning (ICML)}, pages 89--96, 2005.

\bibitem[Chen et~al.(2024)Chen, Xiao, Zhang, Luo, Lian, and Liu]{chen2024bgem3}
Jianlv Chen, Shitao Xiao, Peitian Zhang, Kun Luo, Defu Lian, and Zheng Liu.
\newblock {M3-Embedding}: Multi-linguality, multi-functionality,
  multi-granularity text embeddings through self-knowledge distillation.
\newblock arXiv preprint, 2024.
\newblock arXiv:2402.03216.

\bibitem[{convaiinnovations}(2026)]{hf_laya}
{convaiinnovations}.
\newblock {laya-typed-decisions}.
\newblock Hugging Face,
  \url{https://huggingface.co/convaiinnovations/laya-typed-decisions}, 2026.

\bibitem[Field and Welsh(2007)]{field2007bootstrap}
Christopher~A. Field and Alan~H. Welsh.
\newblock Bootstrapping clustered data.
\newblock \emph{Journal of the Royal Statistical Society: Series B},
  69\penalty0 (3):\penalty0 369--390, 2007.

\bibitem[Gao et~al.(2023)Gao, Yen, Yu, and Chen]{gao2023alce}
Tianyu Gao, Howard Yen, Jiatong Yu, and Danqi Chen.
\newblock Enabling large language models to generate text with citations.
\newblock In \emph{Proceedings of EMNLP}, 2023.
\newblock arXiv:2305.14627.

\bibitem[{GEO-Optim}(2024)]{geobench}
{GEO-Optim}.
\newblock {GEO-Bench} dataset.
\newblock HuggingFace \texttt{GEO-Optim/geo-bench}, 2024.

\bibitem[Hinton et~al.(2015)Hinton, Vinyals, and Dean]{hinton2015distilling}
Geoffrey Hinton, Oriol Vinyals, and Jeff Dean.
\newblock Distilling the knowledge in a neural network.
\newblock arXiv preprint, 2015.
\newblock arXiv:1503.02531.

\bibitem[Hu et~al.(2022)Hu, Shen, Wallis, Allen-Zhu, Li, Wang, Wang, and
  Chen]{hu2022lora}
Edward~J. Hu, Yelong Shen, Phillip Wallis, Zeyuan Allen-Zhu, Yuanzhi Li, Shean
  Wang, Lu~Wang, and Weizhu Chen.
\newblock {LoRA}: Low-rank adaptation of large language models.
\newblock In \emph{International Conference on Learning Representations
  (ICLR)}, 2022.
\newblock arXiv:2106.09685.

\bibitem[{jaredpalmer}(2026)]{hf_kev4b}
{jaredpalmer}.
\newblock {kev-4b}.
\newblock Hugging Face, \url{https://huggingface.co/jaredpalmer/kev-4b}, 2026.

\bibitem[Lakens(2017)]{lakens2017equivalence}
Dani{\"e}l Lakens.
\newblock Equivalence tests: A practical primer for $t$ tests, correlations,
  and meta-analyses.
\newblock \emph{Social Psychological and Personality Science}, 8\penalty0
  (4):\penalty0 355--362, 2017.

\bibitem[Liu et~al.(2023)Liu, Zhang, and Liang]{liu2023verifiability}
Nelson~F. Liu, Tianyi Zhang, and Percy Liang.
\newblock Evaluating verifiability in generative search engines.
\newblock In \emph{Findings of EMNLP}, 2023.
\newblock arXiv:2304.09848.

\bibitem[Liu et~al.(2024)Liu, Lin, Hewitt, Paranjape, Bevilacqua, Petroni, and
  Liang]{liu2024lost}
Nelson~F. Liu, Kevin Lin, John Hewitt, Ashwin Paranjape, Michele Bevilacqua,
  Fabio Petroni, and Percy Liang.
\newblock Lost in the middle: How language models use long contexts.
\newblock \emph{Transactions of the Association for Computational Linguistics},
  12:\penalty0 157--173, 2024.
\newblock arXiv:2307.03172.

\bibitem[{manjunathshiva}(2026)]{hf_opendecider}
{manjunathshiva}.
\newblock {opendecider-small-td}.
\newblock Hugging Face,
  \url{https://huggingface.co/manjunathshiva/opendecider-small-td}, 2026.

\bibitem[{Mapika}(2026)]{hf_decider4b}
{Mapika}.
\newblock {decider-4b}.
\newblock Hugging Face, \url{https://huggingface.co/Mapika/decider-4b}, 2026.

\bibitem[Nogueira et~al.(2020)Nogueira, Jiang, and Lin]{nogueira2020monot5}
Rodrigo Nogueira, Zhiying Jiang, and Jimmy Lin.
\newblock Document ranking with a pretrained sequence-to-sequence model.
\newblock In \emph{Findings of the Association for Computational Linguistics:
  EMNLP 2020}, 2020.
\newblock arXiv:2003.06713.

\bibitem[{Perplexity AI}(2026)]{hf_pplx}
{Perplexity AI}.
\newblock {pplx-decider-v1.1-27b}.
\newblock Hugging Face,
  \url{https://huggingface.co/perplexity-ai/pplx-decider-v1.1-27b}, 2026.

\bibitem[Pfrommer et~al.(2024)Pfrommer, Bai, Gautam, and
  Sojoudi]{pfrommer2024ranking}
Samuel Pfrommer, Yatong Bai, Tanmay Gautam, and Somayeh Sojoudi.
\newblock Ranking manipulation for conversational search engines.
\newblock In \emph{Proceedings of EMNLP}, 2024.
\newblock arXiv:2406.03589.

\bibitem[Pradeep et~al.(2023{\natexlab{a}})Pradeep, Sharifymoghaddam, and
  Lin]{pradeep2023rankvicuna}
Ronak Pradeep, Sahel Sharifymoghaddam, and Jimmy Lin.
\newblock {RankVicuna}: Zero-shot listwise document reranking with open-source
  large language models.
\newblock arXiv preprint, 2023{\natexlab{a}}.
\newblock arXiv:2309.15088.

\bibitem[Pradeep et~al.(2023{\natexlab{b}})Pradeep, Sharifymoghaddam, and
  Lin]{pradeep2023rankzephyr}
Ronak Pradeep, Sahel Sharifymoghaddam, and Jimmy Lin.
\newblock {RankZephyr}: Effective and robust zero-shot listwise reranking is a
  breeze!
\newblock arXiv preprint, 2023{\natexlab{b}}.
\newblock arXiv:2312.02724.

\bibitem[Puerto et~al.(2025)Puerto, Gubri, Green, Oh, and Yun]{puerto2025cseo}
Haritz Puerto, Martin Gubri, Tommaso Green, Seong~Joon Oh, and Sangdoo Yun.
\newblock {C-SEO Bench}: Does conversational {SEO} work?
\newblock In \emph{NeurIPS Datasets and Benchmarks}, 2025.
\newblock arXiv:2506.11097.

\bibitem[Shea and Roche(2026)]{shea2026jev}
Daniel Shea and Se{\'a}n Roche.
\newblock Jev-as-a-judge for agent evals.
\newblock LangChain blog,
  \url{https://www.langchain.com/blog/jev-agent-evals-langsmith}, 2026.
\newblock Published September 2026, accessed September 20, 2026.

\bibitem[Sun et~al.(2023)Sun, Yan, Ma, Wang, Ren, Chen, Yin, and
  Ren]{sun2023rankgpt}
Weiwei Sun, Lingyong Yan, Xinyu Ma, Shuaiqiang Wang, Pengjie Ren, Zhumin Chen,
  Dawei Yin, and Zhaochun Ren.
\newblock Is {ChatGPT} good at search? investigating large language models as
  re-ranking agents.
\newblock In \emph{Proceedings of the 2023 Conference on Empirical Methods in
  Natural Language Processing (EMNLP)}, 2023.
\newblock arXiv:2304.09542.

\bibitem[Tian et~al.(2026)Tian, Chen, and Tang]{tian2026agentgeo}
Zhihua Tian, Yuhan Chen, and Yao Tang.
\newblock Diagnosing and repairing citation failures in generative engine
  optimization.
\newblock arXiv preprint, 2026.
\newblock arXiv:2603.09296.

\bibitem[{TokenRhythm}(2026)]{hf_neohorse}
{TokenRhythm}.
\newblock {NeoHorse-Jev-4B}.
\newblock Hugging Face,
  \url{https://huggingface.co/TokenRhythm/NeoHorse-Jev-4B}, 2026.

\bibitem[{TypeSafe}(2026)]{typesafe2026jev}
{TypeSafe}.
\newblock Jev 1.13 (snapshot 20260917), structured decision model.
\newblock OpenRouter, \url{https://openrouter.ai/typesafe/jev-1.13-20260917},
  2026.
\newblock Released September 18, 2026.

\bibitem[Vishwakarma et~al.(2026)Vishwakarma, Kumar, and
  Jamidar]{vishwakarma2026whatgetscited}
Rahul Vishwakarma, Shushant Kumar, and Ratnesh Jamidar.
\newblock What gets cited: Competitive {GEO} in {AI} answer engines.
\newblock arXiv preprint, 2026.
\newblock arXiv:2605.25517.

\bibitem[Wang et~al.(2023)Wang, Li, Chen, Cai, Zhu, Lin, Cao, Liu, Liu, and
  Sui]{wang2023fair}
Peiyi Wang, Lei Li, Liang Chen, Zefan Cai, Dawei Zhu, Binghuai Lin, Yunbo Cao,
  Qi~Liu, Tianyu Liu, and Zhifang Sui.
\newblock Large language models are not fair evaluators.
\newblock arXiv preprint, 2023.
\newblock arXiv:2305.17926.

\bibitem[Yang et~al.(2025)]{yang2025qwen3}
An~Yang et~al.
\newblock {Qwen3} technical report.
\newblock arXiv preprint, 2025.
\newblock arXiv:2505.09388.

\bibitem[Zhang et~al.(2026)Zhang, He, and Yao]{zhang2026absorption}
Kai Zhang, Xinyue He, and Jingang Yao.
\newblock From citation selection to citation absorption: A measurement
  framework for generative engine optimization across {AI} search platforms.
\newblock arXiv preprint, 2026.
\newblock arXiv:2604.25707.

\bibitem[Zhang et~al.(2025)Zhang, Li, Long, Zhang, Lin, Yang, Xie, Yang, Liu,
  Lin, Huang, and Zhou]{zhang2025qwen3embedding}
Yanzhao Zhang, Mingxin Li, Dingkun Long, Xin Zhang, Huan Lin, Baosong Yang,
  Pengjun Xie, An~Yang, Dayiheng Liu, Junyang Lin, Fei Huang, and Jingren Zhou.
\newblock {Qwen3 Embedding}: Advancing text embedding and reranking through
  foundation models.
\newblock arXiv preprint, 2025.
\newblock arXiv:2506.05176.

\bibitem[Zheng et~al.(2023)Zheng, Chiang, Sheng, Zhuang, Wu, Zhuang, Lin, Li,
  Li, Xing, Zhang, Gonzalez, and Stoica]{zheng2023judge}
Lianmin Zheng, Wei-Lin Chiang, Ying Sheng, Siyuan Zhuang, Zhanghao Wu, Yonghao
  Zhuang, Zi~Lin, Zhuohan Li, Dacheng Li, Eric~P. Xing, Hao Zhang, Joseph~E.
  Gonzalez, and Ion Stoica.
\newblock Judging {LLM}-as-a-judge with {MT-Bench} and {Chatbot Arena}.
\newblock In \emph{Advances in Neural Information Processing Systems (NeurIPS),
  Datasets and Benchmarks Track}, 2023.

\end{thebibliography}

\appendix

\section{Engines}
\label{app:engines}

\begin{table}[h]
\centering
\footnotesize
\setlength{\tabcolsep}{2.5pt}
\resizebox{\linewidth}{!}{%
\begin{tabular}{@{}lllcccccccc@{}}
\toprule
Engine & Vendor & Weights & Queries & Groups & Zero labels & Ranker 4B & Jev & JEV-27B & NIM \\
\midrule
\texttt{mistral-medium-3.5-128b} & Mistral AI & open & 229 & 228 & 22\% & 0.538 & 0.571 & 0.574 & 0.395 \\
\texttt{gemini-3.1-flash-lite} & Google & closed & 150 & 149 & 20\% & 0.599 & 0.650 & 0.621 & 0.425 \\
\texttt{gpt-5.5} & OpenAI & closed & 150 & 143 & 28\% & 0.533 & 0.547 & 0.552 & 0.340 \\
\texttt{gpt-5.4} & OpenAI & closed & 150 & 149 & 24\% & 0.537 & 0.579 & 0.576 & 0.384 \\
\texttt{gpt-5.4-mini} & OpenAI & closed & 150 & 149 & 28\% & 0.556 & 0.544 & 0.557 & 0.405 \\
\texttt{gemma-4-31b-it} & Google & open & 100 & 99 & 23\% & 0.607 & 0.622 & 0.650 & 0.411 \\
\texttt{nex-n2-pro} & Nex AGI & open & 100 & 99 & 28\% & 0.474 & 0.577 & 0.572 & 0.341 \\
\texttt{gpt-oss-120b} & OpenAI & open & 80 & 44 & 67\% & 0.464 & 0.548 & 0.537 & 0.315 \\
\texttt{qwen3.5-122b-a10b} & Alibaba & open & 80 & 80 & 24\% & 0.586 & 0.620 & 0.606 & 0.445 \\
\texttt{llama-3.3-70b-instruct} & Meta & open & 80 & 80 & 17\% & 0.446 & 0.461 & 0.517 & 0.343 \\
 
\bottomrule
\end{tabular}}
\caption{Engine configurations. Queries answered, non-degenerate groups, share of sources with a
zero citation share, and mean within-group Spearman of four scorers per engine; Ranker 4B is GEO
Citation Ranker 4B.}
\label{tab:engines}
\end{table}

The 4B student ranges from 0.446 on \texttt{llama-3.3-70b-instruct} to 0.607 on
\texttt{gemma-4-31b-it}, Jev from 0.461 to 0.650, and the student stays above the NIM reranker on
every engine. \texttt{gpt-oss-120b} gives a zero share to 67\% of sources, which leaves 44
non-degenerate groups out of 80 queries. Averaged with equal weight per engine, the 4B student
scores 0.534, Jev 0.572 and JEV-27B 0.576.

\section{Prompts and rubrics}
\label{app:prompts}

\paragraph{Decision models.} The state lists the query and the passages, each truncated to 250
words. The question for passage $k$ reads: ``Considering the QUERY and comparing PASSAGE $k$
against the four other passages, rate the share of citation PASSAGE $k$ would receive if a
generative answer engine wrote an answer to the query drawing on these 5 sources. Judge RELATIVE
to the other passages. Direct relevance to the query dominates; then citable facts, statistics,
quotability, apparent authority, and self-containment. Do not reward length by itself.'' The six
levels are 0, not cited at all; 1, marginal; 2, minor source; 3, moderate source; 4, major source;
and 5, dominant source, each with a one-line description in the code. The pointwise
protocol asks the same question about a single passage.

\paragraph{LLM judge.} The judge receives a query and five passages in a seeded random order,
truncated to 250 words, and returns a score from 0 to 100 per passage for the expected citation
share relative to the other four. The rubric ranks, in order, direct relevance to the query,
density of citable information, figures and dates, quotability, apparent authority and
self-containment. It instructs the judge not to reward length, to give less to a passage redundant
with a better one, and to ignore presentation order.

\paragraph{Students.} The input is ``Given a user query, judge whether this passage would be cited
by a generative answer engine (Perplexity, ChatGPT Search, Gemini) writing an answer to the
query'', followed by the query and the passage, in the Qwen3-Reranker template. The score is the
log-odds of the ``yes'' token against the ``no'' token.

\end{document}